\documentclass[a4paper]{article}
\usepackage{amsmath,graphicx}
\usepackage{multirow}
\usepackage{graphicx}   
\usepackage{makecell}
\usepackage{subfigure}
\usepackage[para]{footmisc}
\usepackage{INTERSPEECH2021}
\usepackage{color}
\usepackage[para]{footmisc}

\newcommand\blfootnote[1]{%
\begingroup
\renewcommand\thefootnote{}\footnote{#1}%
\addtocounter{footnote}{-1}%
\endgroup
}

\title{E2E-based Multi-task Learning Approach 
to Joint Speech and Accent Recognition}
\name{Jicheng Zhang$^1$, Yizhou Peng$^1$, Pham Van Tung$^2$, Haihua Xu$^2$, Hao Huang$^1$, Eng Siong Chng$^2$}
\address{
  $^1$School of Information Science and Engineering, Xinjiang University, Urumqi, China\\
  $^2$School of Computer Science and Engineering, Nanyang Technological University, Singapore}
\email{}
\begin{document}

\maketitle
\begin{abstract}
In this paper, we propose a single multi-task learning framework to perform End-to-End (E2E) speech recognition (ASR) and accent recognition (AR) simultaneously.
The proposed framework is not only more compact but  can also yield comparable or even better results than standalone systems.
Specifically, we found that the overall performance is predominantly determined by the ASR task, and the E2E-based ASR pretraining is essential to achieve improved performance, particularly  for the AR task.
Additionally, we conduct several analyses of the proposed method.
First, though the objective loss for the AR task is much smaller compared with its counterpart of ASR task, a smaller weighting factor with the AR task in the joint objective function is necessary to yield better results for each task.
Second,
we found that sharing only a few layers of the encoder yields better AR results than sharing the overall encoder.
Experimentally, the proposed method produces WER results close to the best standalone E2E ASR ones, while it achieves 7.7\% and 4.2\% relative improvement over standalone and single-task-based joint recognition methods on test set for accent recognition respectively.

\end{abstract}
\section{Introduction}\label{sed:introduction}
The advent of End-to-end (E2E) modeling approach in machine learning  has  fundamentally changed speech community in recent years. The main success of the E2E method lies in its simplicity, compactness, as well as effectiveness. Take Automatic Speech Recognition (ASR)  for example, we boil down conventional acoustic, language, as well as lexicon modeling work into an entire network
that typically contains two components, that is, the encoder and decoder.  Both encoder and decoder are concatenated with attention sub-network and the whole network  is optimized with the same objective functions. 
\blfootnote{~~Students ($^*$) have joined SCSE MICL Lab, NTU, Singapore as exchange students.
This work is supported by the National Key R\&D Program of China (2017YFB1402101), Natural Science Foundation of China (61663044, 61761041), Hao  Huang is the corresponding  author.}

Although very successfully applied to wide range of speech processing areas, the potential of E2E modeling framework has yet to be fully unleashed. For instance, one  can employ E2E framework to perform diversified speech tasks, such as speaker recognition~\cite{arasteh2020generalized,shon2018frame}, accent recognition~\cite{huang2021aispeech}, as well as speech recognition~\cite{kubo2020joint,zhang2020unified,winata2020learning}, just to name a few, aimed at a single task (mono-task) at a time. Intuitively, this is different to what humans are doing.
To recognize an incoming speech signal, human can easily perform speech recognition, as well as other recognition, such as speaker identity, gender, and accent recognition etc. simultaneously. In other words, humans are multi-task recognition performer, while present E2E framework are mainly targeted at mono-task recognition, and its multi-task capability is not fully demonstrated.

In this paper, we attempt an E2E-based multi-task learning approach to conduct joint speech and accent recognition simultaneously. 
Our work is motivated by a workshop participation~\cite{shi2021accented}. 
There are  two tracks of the challenge, that is, 8 accented English speech recognition and corresponding accent classification respectively. 
Under the E2E modeling framework, both tasks can be easily merged into a single architecture to perform~\cite{viglino2019end}. 
We are curious if we can still 
achieve state-of-the-art results on each task with a single multi-task modeling framework. 
It is not only simpler but 
is also close to what our humans are doing, namely, performing speech and accent recognition simultaneously.

The main contributions of the work lie in the following aspects. 1) By multi-task learning, we mean our focus is to use a single network architecture to undertake  two tasks simultaneously. 
This differentiates our work from the prior works, since majority of them only focus on accented ASR with multi-task learning~\cite{jain2019multi,zhu2020multi,yang2018joint,rao2017multi,ghorbani2019leveraging,jain2018improved}. 
2) We found ASR pretraining (with or without out-domain data)  is essential to achieve better results on either task, particularly for the accent recognition task~\cite{huang2020cross}. 
3) By scaling down the accent recognition loss in our combined objective function during training, we can obtain better results on either task consistently.

The rest of this paper is organized as follows. Section~\ref{sec:related-work} describes related works. Section~\ref{sec:model-architerture} describes the model architecture of the proposed multi-task method and corresponding optimization method. Section~\ref{sec:data-desp}
 describes the overall data for experiments, and  Section~\ref{sec:exp-setup} briefs the experimental setups. Section~\ref{sec:res} reports the experimental results, and it is followed by the analysis in Section~\ref{sec:ana}. Finally, we conclude in Section~\ref{sec:con}.

\section{Related work} \label{sec:related-work}
Accented ASR~\cite{ghorbani2018advancing,2019Multi} and accent 
recognition~\cite{berkling1998improving,jiao2016accent} have been extensively studied recently, but majority of the research works on the two regards are either performed separately or using one to boost another with multi-task method~\cite{rao2017multi,ghorbani2019leveraging,jain2018improved,li2018multi}.  Assuming an  accented English is a product of foreign English influenced by a specific native language, \cite{ghorbani2019leveraging} proposed  a multi-task learning approach to train an improved accented speech recognition model.
Similarly with the multi-task framework, \cite{rao2017multi} employed accented phoneme-based multi-task learning to boost accented ASR performance, while both \cite{jain2018improved} and \cite{li2018multi} proposed to use accent embedding for improved accented speech recognition.
Aside from multi-task learning approach,
\cite{jain2019multi} proposed a learning-based mixture of  transform dealing with phone  and  accent variability in a joint framework.
For accent recognition, the work in \cite{huang2021aispeech} has employed phone posteriorgram (PPG) features as input to train the accent classifier for the accent challenge track. With diversified data augmentation, as well as system fusion methods, it has achieved the best results out of scores of participants.

From the above-mentioned, 
one of the most closely related work to us is ~\cite{viglino2019end}, 
which not only employed multi-task learning approach to speech and accent recognition, but also reported the performance for each task, though their main focus is for improved accented speech recognition.
The main distinctions are two aspects. First, the contribution of ASR pretraining are not studied in ~\cite{viglino2019end,ghorbani2018advancing,2019Multi}, particularly for the efficacy on accent recognition. 
Second, the objective losses from the two tasks are not analyzed.
We found though the two losses are not on the same range, and ASR loss is much bigger.
Therefore, scaling down the accent objective loss consistently yields better results for each recognition task.

\section{Model architecture and optimization}\label{sec:model-architerture}
\subsection{Architecture}\label{sub:architecture}
\begin{figure} [ht]
    \centering
     \includegraphics[width=8cm]{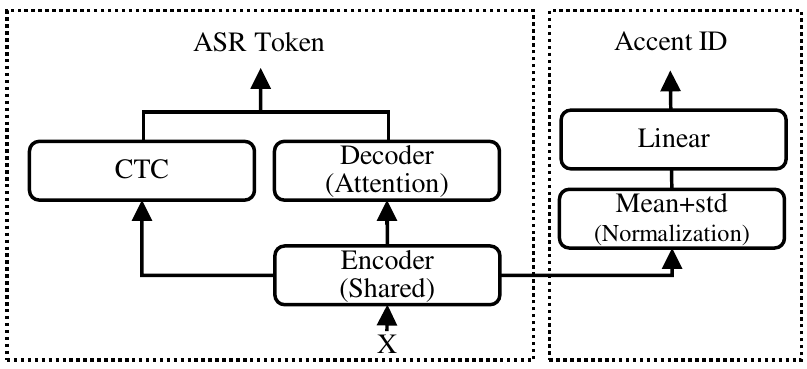}
    \caption{Multi-task network architecture for joint End-to-End speech recognition and accent recognition.
    }
    \label{fig:architecture}
\end{figure}

In this paper, all of our work is performed with Transformer-based E2E framework~\cite{dong2018speech}, from the baseline standalone ASR, accent recognition, to the proposed 
Multi-Task-based Joint Recognition (MTJR).
Figure~\ref{fig:architecture} illustrates the entire network architecture for the proposed method.
\setlength{\belowcaptionskip}{-0.05cm}

From Figure~\ref{fig:architecture}, our ASR network is a standard Transformer-based E2E network~\cite{kim2017joint}, including
encoder, decoder components boosted with self-attention, as well as cross-attention mechanism.
We note that the output of the ASR are independent of accent.
For the accent recognition network (on the right-hand side of Figure~\ref{fig:architecture}), it has a shared encoder with the ASR network. With the outputs from the shared encoder, a pooling layer is employed to estimate the mean and variance vectors.
We then concatenate the mean and variance vectors as input to a linear transform followed by a softmax layer to estimate each accent probability.
\subsection{Optimization}\label{sub:optimization}
From Figure~\ref{fig:architecture}, the Encoder is jointly optimized with two objective losses, one is from ASR, denoted as $\mathcal{L}_{asr\underline{~~}{loss}}$, while the other is from accent recognition, denoted as $\mathcal{L}_{{accent}\underline{~~}{loss}}$; that is, the total objective loss is as Equation~\ref{eqn:multi_loss}.
 \begin{equation}
  \mathcal{L}_{loss}={\beta\mathcal{L}}_{asr\underline{~~}loss}+{\lambda\mathcal{L}}_{accent\underline{~~}loss}
  \label{eqn:multi_loss}
\end{equation}
where both $\beta$ and $\lambda$ are hyper-parameters, and normally they are subject to $\beta+\lambda=1$. To simplify the tuning process, we train the multi-task network with  Equation~\ref{eqn:simple-multi-loss} in practice.
 \begin{equation}
  \mathcal{L}_{loss}={\mathcal{L}}_{asr\underline{~~}loss}+{\lambda\mathcal{L}}_{accent\underline{~~}loss}
  \label{eqn:simple-multi-loss}
\end{equation}
where we fix the ASR loss, while keep tuning the loss from accent recognition. 
This gives us more flexibility to investigate how the overall performance is impacted by the accent recognition task.
We will analyze this in Section~\ref{sub:tune-lambda}  using Equation~\ref{eqn:simple-multi-loss} to optimize.
Meanwhile, the $\mathcal{L}_{asr\underline{~~}{loss}}$ in Equation~\ref{eqn:simple-multi-loss} is realized as Equation~\ref{eqn:asr}.
\begin{equation}
  \mathcal{L}_{asr\underline{~~}loss}={\gamma\mathcal{L}}_{ctc}+{(1-\gamma)\mathcal{L}}_{att}
  \label{eqn:asr}
\end{equation}
That is, our ASR network is jointly optimized with connectionist temporal classification~ (CTC)~\cite{graves2006connectionist} and attention network objective functions, and 
$\gamma$ is also a hyper-parameter.
In all our following experiments, we found that $\gamma$ = 0.3 can get better results.
\section{Data description}\label{sec:data-desp}
For the accented ASR and accent recognition challenges in the workshop~\cite{shi2021accented}, the organizer allows each participant to use two categories of data, i.e.,  in-domain and out-domain data.
The in-domain data is sponsored by \texttt{DataTang} Company in China\footnote{https://www.datatang.com/INTERSPEECH2020}.
It contains 8 accents of English read speech data, that are,
American(US), British(UK), Chinese(CHN), Indian(IND), Japanese(JPN), Korean(KR), Portuguese(PT) and Russian(RU) accent respectively. Each accent has roughly 20 hours. The out-domain data is Librispeech with 960 hours, which is optional for participants to use.
Besides, there is no accent information available for Librispeech data. 
Table~\ref{tab:data-spec} describes the overall data distributions.
\begin{table}[htbp]
 \centering
 \caption{Data description for both accented ASR and accent recognition challenges. The so-called ``in-domain" part refers to the required data sets for each participant, while ``out-domain" data is optional to use. }
 \label{tab:data-spec}
\begin{tabular}{c|c|c|c|c}
 \hline
   Category & Name & Len (Hrs) & Word/Utt. & Sec./Utt. \\
   \hline
   \multirow{3}*{\makecell{In-domain}} 
   & Train & 148.5 & 9.72 & 4.29\\
   & Dev& 14.5 & 9.66 & 4.35 \\
   & Test& 20.95 & 9 & 4.15 \\
   \hline
  Out-domain & Libri. & 961 & 33.43 & 12.3 \\
    \hline
\end{tabular}
\end{table}
\begin{table*}[htbp]
 \centering
 \caption{Accent recognition accuracy (\%) with different Transformer configurations, with or without ASR pretraining using in-domain or out-domain training data sets over \texttt{Test} data set in Table~\ref{tab:data-spec}.}
 \label{tab:mono-accent-res}
\begin{tabular}{c|c|c|c c c c c c c c}
 \hline
    & System & Avg & US & UK & CHN & IND & JPN & KR & PT & RU \\
  \hline
    Transformer-3L & S1 & 30.7 & 21.1 & 56.9 & 31.3 & 41.6 & 31.2 & 9.7 & 41.1 & 18.1 \\ 
Transformer-6L & S2 & 29.7 & 15.9 & 44.2 & 32.4 & 33.6 & 21.4 & 24.6 & 35 & 36 \\ 
Transformer-12L & S3 & 32.6 & 20.8 & 40.6 & 22.7 & 46.2 & 38.8 & 36.7 & 35.2 & 23.8 \\ 
    ASR-pretrain + S1 & S4 & 45 & 49.7 & 73.5 & 20.1 & 54.2 & 45 & 41.3 & 43.4 & 36.1 \\ 
    ASR-pretrain + S2 & S5 & 64.8 & 55.6 & 81.4 & 57.9 & 82 & 53.6 & 64.1 & 63.3 & 65.5 \\ 
    ASR-pretrain + S3 & S6 & 66.6 & 55.4 & 83.9 & 67.5 & 83.9 & 49.3 & 76.7 & 65 & 55.4 \\ 
    S6 + Libri. & S7 & 69.8 & 52 & 92 & 73 & 86.4 & 55.1 & 78 & 70.3 & 57.8 \\
    \hline
\end{tabular}
\end{table*}
\section{Experimental setup}\label{sec:exp-setup}
All experiments are performed using E2E-based Transformer modeling framework with Espnet tookit~\cite{watanabe2018espnet}.
For comparison, we train two groups of models, one is mono-task models that are either for speech recognition or for accent recognition only, and the other is joint models that are realizing two kinds of approaches to conduct joint speech and accent recognition simultaneously.
Specifically, one follows the work in~\cite{hou2020large,watanabe2017language}, 
which is actually a normal Transformer ASR system but appending each decoded utterance with a recognized accent label, 
we call it single-task-based joint recognition~(STJR)  method, and the other is our proposed MTJR method.
For speech and accent recognition tasks, the encoders of the Transformer method are the same, while the decoders are different, and it is much simpler in the case of accent recognition task as mentioned in Section~\ref{sub:architecture}.

We have conducted two kinds of experiments to train the models mentioned above, one is to train them from scratch, and the other is to employ pretraining. 
For pretraining, we use either in-domain data only, or the overall data that includes both in-domain and out-domain data, to train a Transformer-based ASR system. 
Then we fine-tune the initialized components with in-domain data for the two tasks respectively.

For the actual training,  we use the acoustic feature that is 83-dimensional with 80-d filter bank plus 3-d pitch features~\cite{ghahremani2014pitch}.
The Transformer ASR and the two joint methods are configured with 12-layer encoder, 6-layer decoder with 4-head attention, while
the mono-task-based accent recognition method uses three different numbers of layers~(3/6/12) in encoder for experimentation.
During training, we proceed with 0.1 drop-out, and the Transformers are optimized with Noam optimizer~\cite{vaswani2017attention}. The output of ASR has 2000 BPEs~\cite{sennrich2015neural}. Besides,
to yield robust models, we adopt 
various data augmentation methods including spec-augmentation~\cite{park2019specaugment}, as well as speed perturbation~\cite{ko-2015-speed_perturb}.

\section{Results}
\label{sec:res}
Table~\ref{tab:mono-accent-res} reports our mono-task-based accent recognition results using Transformer methods with or without ASR pretraining over \texttt{Test} data set in Table~\ref{tab:data-spec}. 
From Table~\ref{tab:mono-accent-res}, we observe bigger Encoder (12-layer) model produces better results. 
Besides, we obtain significantly better classification results while employing ASR pretraining method. 
Moreover, it benefits further when using out-domain data (Librispeech) to pretrain the model. 
Comparing the results between systems with or without pretraining, we observer the best average accuracy (69.8\%) of the pretrained system (S7) is more than double of that (32.6\%) of the best system (S3) without pretraining at all. 
Notably, we also get similar results 
on \texttt{Dev} data set, however we skip it due to space limitation.

Table~\ref{tab:joint-res} reports the results of both joint training and multi-task learning with (M2, M4, M5, M7, M8) or without (M1, M3, M6) ASR pretraining. 
For the two joint learning methods, we employ both in-domain-only and the overall (Librispeech included) data based pretraining respectively. To perform the recipe, we first use the data to train a normal Transformer ASR system as indicated in the left-hand side of Figure~\ref{fig:architecture}.
After that we change the output embedding and projection of the Transformer for the target system and use the in-domain data to fine-tune the system.

For the systems (M1, M3, and M6) without pretraining recipe in Table~\ref{tab:joint-res},  we can see the proposed MTJR method achieves both better ASR and better accent recognition results on both \texttt{Dev} and \texttt{Test} data sets, compared to the STJR method proposed in~\cite{hou2020large,watanabe2017language}, that is, M3 versus M6, in Table~\ref{tab:joint-res}. 
However, the ASR results from both joint recognition methods are worse than those of the mono-task ASR, namely, M3 and M6 versus M1 respectively in Table~\ref{tab:joint-res}. 
This suggests the accent recognition task has some negative effects on the ASR performance.

For the systems (M2, M4, M5, M7 and M8)  with pretraining method in Table~\ref{tab:joint-res}, we achieve significant ASR performance improvement in every case. What we observe here is agreed with the work in \cite{huang2020cross} which also observed ASR pretraining is essential to improve the results for either mono-lingual, multi-lingual or cross-lingual task.
For the ASR results of the 3 methods in Table~\ref{tab:joint-res}, the proposed MTJR method results slightly worse than those of the Transformer and STJR method respectively. This again reminds us looking into how accent recognition task affects the ASR performance. 
To put it simply, we conjecture there would be an $\lambda$ in Eq.~\ref{eqn:simple-multi-loss} or other factors leading to optimal results on either tasks for the MTJR case, which  are  to be analyzed  in Section~\ref{sec:ana} specifically.

Regarding to the accent recognition results in Table~\ref{tab:joint-res},
we observe that the proposed MTJR method is not only better than the STJR method without pretraining~(M3 versus M6),
it also benefits significantly more with pretraining method. 
It consistently outperforms the STJR method when we compare M4 versus M7, and M5 versus M8 respectively.
Particularly, MTJR method achieves 75.2\% average accent recognition accuracy while the STJR method only achieves 72.2\% accuracy on the \texttt{Test} set,  when the overall data is used for the pretraining.
Such an accent recognition result places us in the second position out of more than 70 participants~\cite{shi2021accented}, and the best result~\cite{huang2021aispeech} is obtained with 60 times of data augmentation as well as feature-based system fusion. 
For the STJR method as in~\cite{hou2020large,watanabe2017language},
we note that we conducted 2 kinds of joint training recipes, that is, either output a accent label before actual ASR output or append an accent label after ASR final output. We have not observed much WER difference from the two setups. 
\begin{table}[htbp]
    \centering
        \caption{Joint speech and accent recognition results with or without ASR pretraining. All systems use speed perturb and spectral augmentation(specAug) by default. For MTJR, we set $\lambda=0.1$. M2,M5,M8 use all data to train background model, but M4,M7  use in-domain data only.
        }
    \begin{tabular}{c|c|c|c|c|c}
        \hline
        \multirow{2}*{Method} & \multirow{2}*{System} & \multicolumn{2}{c|}{WER(\%)} & \multicolumn{2}{c}{ACC(\%)} \\ \cline{3-4}
        \cline{5-6}
        & & Dev & Test & Dev & Test \\
        \hline
        \multirow{2}*{Transformer} 
            & M1 & 6.9 & 8 & -- & -- \\
            & M2 & 5.7 & 6.7 & -- & -- \\ 
        \hline
        \multirow{3}*{STJR~\cite{hou2020large,watanabe2017language}} 
            & M3 & 9.1 & 10.6 & 80.5 & 69.7 \\ 
            & M4 & 7 & 8 & 75.7 & 68.4 \\ 
            & M5 & 5.8 & 6.6 & 77 & 72.2 \\ 
        \hline
        \multirow{3}*{MTJR} 
            & M6 & 8.1 & 9.5 & 80.2 & 71 \\ 
            & M7 & 7.3  & 8.4 & 81.8 & 72.4 \\ 
            & M8 & 6.2 & 7.1 & 82.4 & 75.2 \\ 
        \hline
    \end{tabular}
    \label{tab:joint-res}
\end{table}
\section{Analysis}~\label{sec:ana}
As observed from Table~\ref{tab:joint-res}, though the proposed MTJR method can achieve much better accent recognition results, 
it's ASR WERs are little bit worse compared to those of the  Transformer or STJR methods. 
Here, we are curious about in what situation we can achieve desired results for either task. 
\subsection{Objective losses}~\label{sub:losses}
\begin{figure}[htbp]
    \centering 
    \begin{tabular}{cc}
        \begin{minipage}[t]{0.49\linewidth}           \includegraphics[width=\linewidth, height=0.6\linewidth]{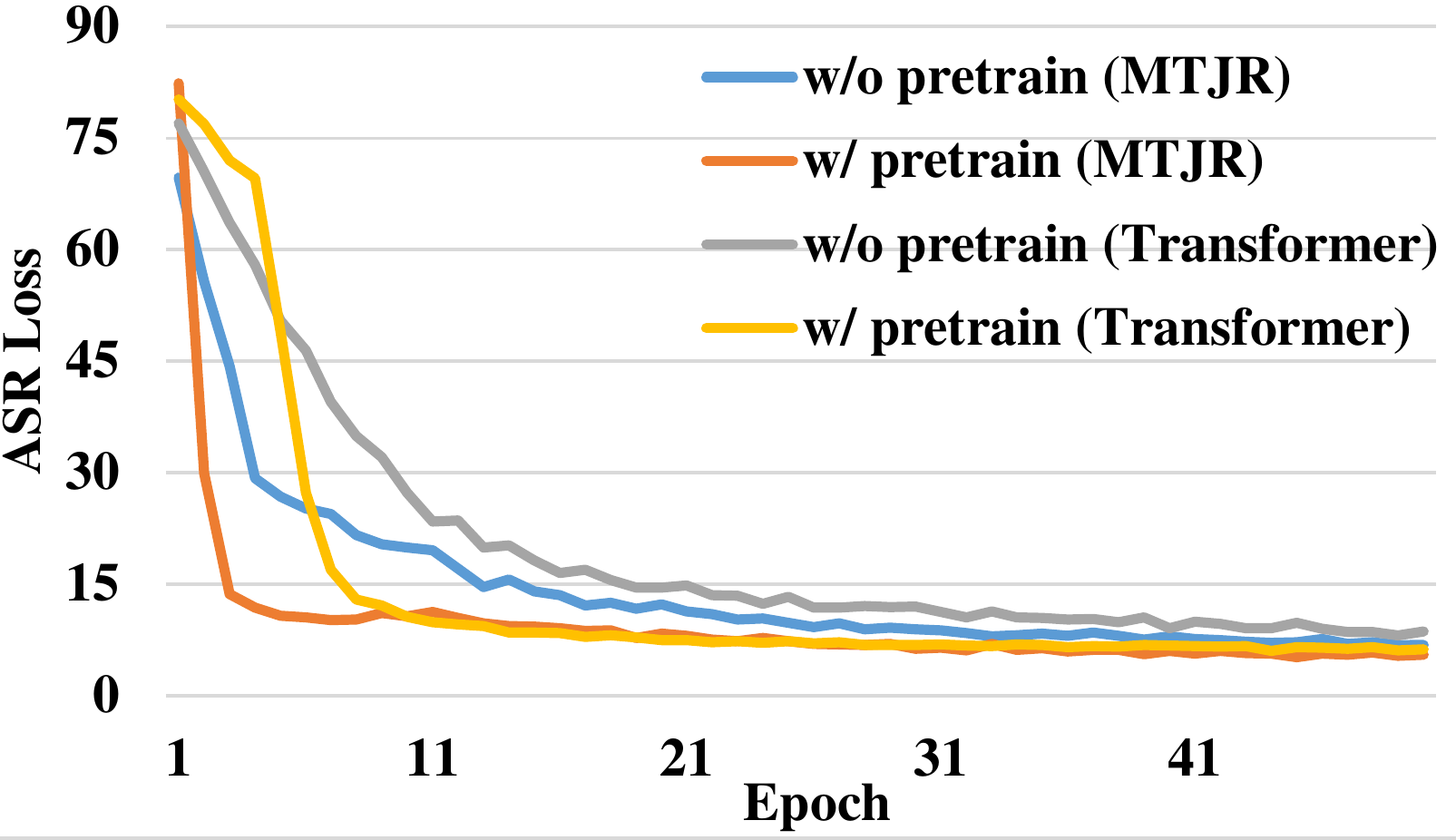}
            \centerline{(a) ASR loss versus epoch}\medskip
        \end{minipage}
        \begin{minipage}[t]{0.49\linewidth}
            \includegraphics[width=\linewidth, height=0.6\linewidth]{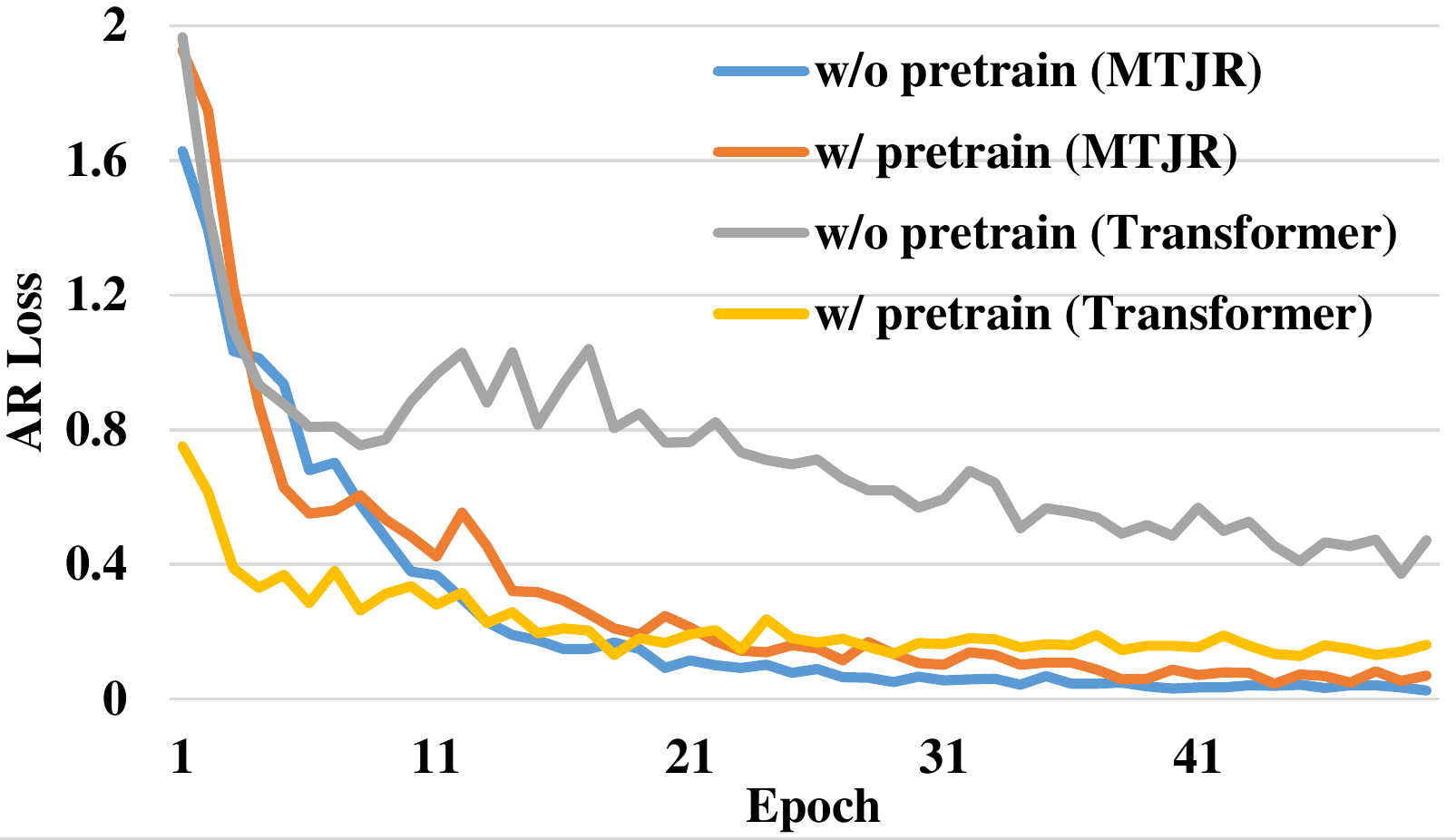}
            \centerline{(b) AR loss versus epoch}\medskip
        \end{minipage}
    \end{tabular}
    \caption{Speech and Accent Recognition  losses versus training epoch for MTJR method.}
    \label{fig:loss-curve}
\end{figure}

To answer the question, it seems that we just simply tune the $\lambda$ in Eq.~\ref{eqn:simple-multi-loss}. However, we want to examine the objective losses versus training epochs  for the two tasks first. 
This gives us more insight about the property of the two tasks. 
Figure~\ref{fig:loss-curve} plots the loss versus training epoch for both speech and accent recognition respectively.

One of the main points worth our attention from Figure~\ref{fig:loss-curve} is the ASR loss value is obviously larger than the accent recognition loss value. 
The latter is so small so as to be negligible compared with the former. Though with small loss value, the accent recognition task still has negative effect on the ASR performance as shown in Table~\ref{tab:joint-res}.
Next, we see how to obtain optimal results by means of $\lambda$ tuning. 
\subsection{Tuning $\lambda$ for optimal results}\label{sub:tune-lambda}
\begin{figure} [ht]
    \centering
     \includegraphics[width=8cm]{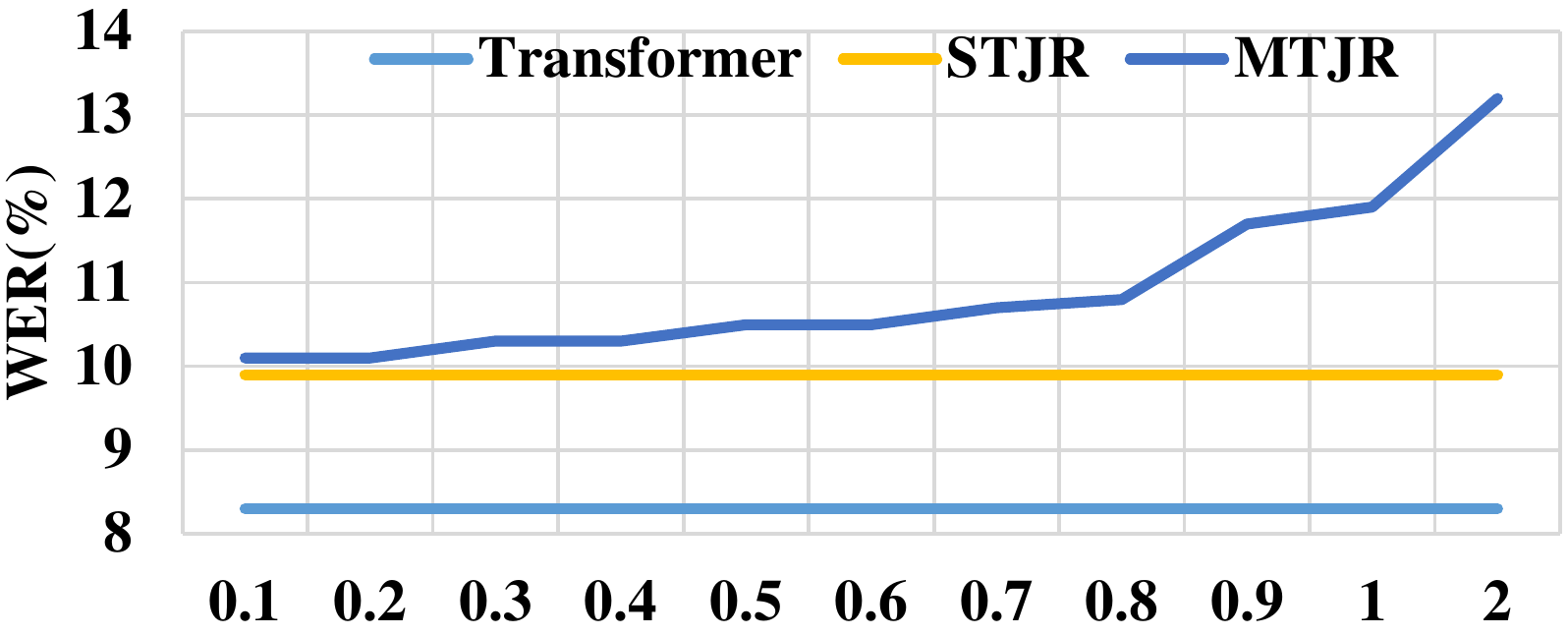}
    \caption{Test set WER versus $\lambda$ for MTJR method.}
    \label{fig:wer-versus-lambda}
\end{figure}

\begin{figure} [ht]
    \centering
     \includegraphics[width=8cm]{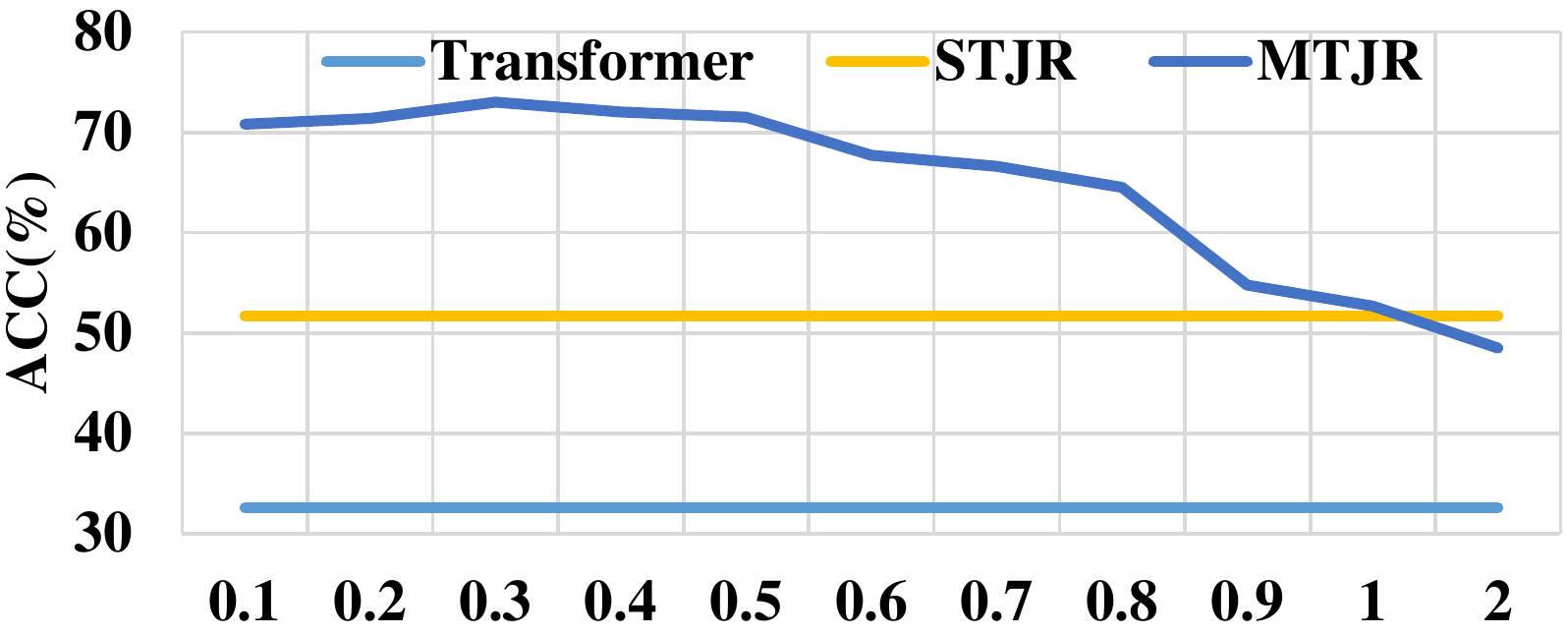}
    \caption{Test set accent recognition accuracy versus $\lambda$ for MTJR method.}
    \label{fig:accent-res-versus-lambda}
\end{figure}

Figures~\ref{fig:wer-versus-lambda} and 
\ref{fig:accent-res-versus-lambda} plot speech and accent recognition results versus $\lambda$ in Eq.~\ref{eqn:simple-multi-loss} on \texttt{Test} set respectively for the MTJR method.

From either Figure, we see that as $\lambda$ increases, either task yields degraded results though the accent loss is rather small as shown in Figure~\ref{fig:loss-curve}. Specifically, we found that keeping $\lambda$ around [0.1, 0.3] can let us have better results on both tasks.
\subsection{Multi-task encoder-layer sharing}\label{sub:encoder-share}
It is a common interpretation that the initial layers of the encoder are expected to encode lower level acoustic information, while subsequent layers codify more complex cues closer to phonetic classes~\cite{bengio2013representation}.
For E2E-based speech recognition, one can reasonably think encoder is responsible for feature learning while decoder is for linguistic interpretation. However, prior work~\cite{fan2020exploring} even shows different layers of encoder are learning for different tasks.
As a result, it naturally comes up with a question: which layer sharing can yield better results for speech or accent recognition. Table~\ref{tab:multi-diff-layer} reports the experimental results with different encoder layers being shared between speech recognition and accent recognition tasks. 

\begin{table}[htbp] 
    \centering
        \caption{The results of MTJR method by different encoder layers sharing with $\lambda=0.1$ in Eq.~\ref{eqn:simple-multi-loss}.}
    \begin{tabular}{c|c|c|c|c|c}
        \hline
        \multirow{2}*{Shared layers} & \multirow{2}*{System} & \multicolumn{2}{c|}{WER(\%)} & \multicolumn{2}{c}{ACC(\%)} \\ \cline{3-4}
        \cline{5-6}
        & & Dev & Test & Dev & Test \\
        \hline
        \multirow{1}*{Layer-3} 
            & L1 & 8.7 & 10 & 64.8 & 54.6 \\
        \hline
        \multirow{1}*{Layer-6} 
            & L2 & 8.6 & 10.1 & 78.3 & 66 \\ 
        \hline
        \multirow{1}*{Layer-9}  
            & L3 & 8.7 & 9.9 & 80.2 & 71.4 \\
        \hline
        \multirow{1}*{Layer-12}  
            & L4 & 8.6 & 10.1 & 77.1 & 70.8 \\
        \hline
    \end{tabular}
    \label{tab:multi-diff-layer}
\end{table}
We see from Table~\ref{tab:multi-diff-layer} the best  recognition results are obtained when 9 layers of the overall 12-layer encoder are shared.
With a fixed $\lambda$, the WER results are not obviously affected by how many layers are shared. However, the accent recognition results are different. 
Only lower layer sharing (L1 and L2 in Table~\ref{tab:multi-diff-layer}) yields much worse results, while overall layer sharing (L4) does not mean the best results. 
We note that results in Table~\ref{tab:multi-diff-layer}  cannot be compared with what are presented in Table~\ref{tab:joint-res}, since speed perturbation is  employed in Table~\ref{tab:joint-res}
while it is not in Table~\ref{tab:multi-diff-layer} for simplicity. 
\section{Conclusions}
\label{sec:con}
In this paper we proposed an E2E-based multi-task learning approach to joint speech and accent recognition. 
The benefit of the method is it can perform speech and accent recognition simultaneously. We verify its effectiveness on an accented English data set that is released for speech and accent recognition contests in a workshop. 
Compared to a conventional mono-task Transformer ASR system, the ASR performance is slightly degraded, while its performance on the accent recognition is noteworthy, especially when the mutli-task system is initialized with ASR pretraining using in-domain and out-domain data. 
Besides, we conducted a series of analysis work. 
We found though the objective loss of the accent recognition task is much smaller, scaling down it in the joint objective function yields better results for both recognition tasks. Finally, we also investigated different encoder-layer sharing for the multi-task method,
and found partial sharing (at 9th layer), instead of the overall layer (12 layers)  of the encoder, produced the best accent recognition results.

\vfill\pagebreak
\clearpage
\bibliographystyle{IEEEtran}

\bibliography{mybib}
\end{document}